 \UseRawInputEncoding
\documentclass[%
 reprint,
superscriptaddress,
floatfix,
 amsmath,amssymb,amsfonts,
prb,
]{revtex4-2}

\usepackage{graphicx}
\usepackage{dcolumn}
\usepackage{bm, bbold}
\usepackage{comment, color}
\usepackage{natbib}
\usepackage{tabularx}
\usepackage{xcolor}
\usepackage{hyperref}

 \usepackage[ngerman, english]{babel}

 \newcommand{\dc}[1]{\begingroup\color{black}#1\endgroup}

\begin{document}


\def\bea{\begin{eqnarray}}
\def\eea{\end{eqnarray}}
\def\beq{\begin{equation}}
\def\eeq{\end{equation}}
\def\f{\frac}
\def\k{\kappa}
\def\e{\epsilon}
\def\ve{\varepsilon}
\def\be{\beta}
\def\D{\Delta}
\def\h{\theta}
\def\t{\tau}
\def\a{\alpha}

\def\cDa{{\cal D}[X]}
\def\cD{{\cal D}[x]}
\def\cL{{\cal L}}
\def\cLo{{\cal L}_0}
\def\cLa{{\cal L}_1}
\def\rv{{\bf r}}
\def\tv{\hat t}
\def\on{{\omega_{\rm a}}}
\def\od{{\omega_{\rm d}}}
\def\off{{\omega_{\rm off}}}
\def\fv{{\bf{f}}}
\def\fm{\bf{f}_m}
\def\bv{{\bf{v}}}
\def\zh{\hat{z}}
\def\yh{\hat{y}}
\def\xh{\hat{x}}
\def\km{k_{m}}
\def\nh{\hat{n}}

\def\Re{{\rm Re}}
\def\sj{\sum_{j=1}^2}
\def\rk{\rho^{ (k) }}
\def\rek{\rho^{ (1) }}
\def\cek{C^{ (1) }}
\def\rz{\rho^{ (0) }}
\def\rt{\rho^{ (2) }}
\def\rtb{\bar \rho^{ (2) }}
\def\trk{\tilde\rho^{ (k) }}
\def\trek{\tilde\rho^{ (1) }}
\def\trz{\tilde\rho^{ (0) }}
\def\trt{\tilde\rho^{ (2) }}
\def\r{\rho}
\def\tD{\tilde {D}}

\def\bt{{\bf t}}
\def\s{\sigma}
\def\kb{k_B}
\def\bF{\bar{\cal F}}
\def\F{{\cal F}}
\def\la{\langle}
\def\ra{\rangle}
\def\nn{\nonumber}
\def\up{\uparrow}
\def\dn{\downarrow}
\def\S{\Sigma}
\def\dg{\dagger}
\def\d{\delta}
\def\p{\partial}
\def\l{\lambda}
\def\L{\Lambda}
\def\G{\Gamma}
\def\o{\Omega}
\def\w{\omega}
\def\g{\gamma}
\def\E{{\mathcal E}}

\def\O{\Omega}

\def\vv{ {\bf v}}
\def\jv{ {\bf j}}
\def\jr{ {\bf j}_r}
\def\jd{ {\bf j}_d}
\def\jdd{ { j}_d}
\def\noi{\noindent}
\def\a{\alpha}
\def\d{\delta}
\def\p{\partial} 

\def\la{\langle}
\def\ra{\rangle}
\def\e{\epsilon}
\def\n{\eta}
\def\g{\gamma}
\def\break#1{\pagebreak \vspace*{#1}}
\def\hf{\frac{1}{2}}
\def\rcurs{r_{ij}}

\def\bv{ {\bf b}}
\def\uv{ {\bf u}}
\def\rv{ {\bf r}}
\def\cf{{\mathcal F}}
\definecolor{MyOrange}{RGB}{255, 140, 0}
\newcommand{\alert}{}

\title{Mechanistic insights into Z-ring formation and stability: A Langevin dynamics approach to FtsZ self-assembly} 
\author{Rajneesh Kumar}
\email{rajneesh.kumar@iopb.res.in}
\affiliation{Institute of Physics, Sachivalaya Marg, Sainik School, Bhubaneswar 751005, India}
\author{Ramanujam Srinivasan}
\email{rsrini@niser.ac.in}
\affiliation{National Institute of Science Education and Research, Khurda 752050, Odisha, India}
\affiliation{Homi Bhabha National Institutes (HBNI), Training School Complex, Anushakti Nagar, Mumbai, India 400094}

\author{Debasish Chaudhuri}
\email[Corresponding author:~]{debc@iopb.res.in}
\affiliation{Institute of Physics, Sachivalaya Marg, Sainik School, Bhubaneswar 751005, India}
\affiliation{Homi Bhabha National Institutes (HBNI), Training School Complex, Anushakti Nagar, Mumbai, India 400094}

\begin{abstract}
The tubulin-like protein FtsZ is crucial for cytokinesis in bacteria and many archaea, forming a ring-shaped structure called the Z-ring at the site of cell division. Despite extensive research, the self-assembly of Z-rings is not entirely understood. We propose a theoretical model based on FtsZ's known filament structures, treating them as semiflexible polymers with specific mechanical properties and lateral inter-segment attraction that can stabilize ring formations. Our molecular dynamics simulations reveal various morphological phases, including open helices, chains, rings, and globules, capturing experimental observations in the fission yeast model using FtsZ from different bacterial species or mutants of {\em Escherichia coli}. Using our theoretical model, we explore how treadmilling activity affects Z-ring stability and identify a spooling mechanism of ring formation. The active ring produces contractile, shear, and rotational stresses, which intensify as the Z-ring transitions to an open helix at high activity. 

\end{abstract}

\maketitle

\section{Introduction} \label{sec:I}
FtsZ is one among the diverse family of bacterial cytoskeletal proteins that exhibits distinct homologies to eukaryotic tubulin~\cite{Bi1991, Adams2009, Harold2010, Lutkenhaus1980}. 
In prokaryotic cells, e.g., bacteria and many archaea, FtsZ (filamenting temperature-sensitive mutant Z) proteins assemble into a ring, the Z-ring, at the future division site, and it attaches to the cell membrane ~\cite{Haeusser2016, Ortiz2015}. This ring provides the backbone for the formation of  divisome~\cite{Vicente2006, Nanninga1991}, a complex multi-protein structure that recruits other proteins required for septum formation between dividing cells. Moreover, the Z-ring itself may generate a constriction force helping cytokinesis~\cite{Osawa2008a}. 

{
Assembly of FtsZ into a ring-like structure requires polymerization of the monomeric units of FtsZ, which, like tubulin, depends on the binding of the nucleotide GTP~\cite{deBoer1992, RayChaudhuri, Mukherjee1994, Lowe1998, Nogales1998}. Although GTP hydrolysis triggers the disassembly of FtsZ filaments, the assembly and polymerization of FtsZ itself are required for the GTPase activity. This is because GTP is sandwiched in the interface between two subunits of FtsZ, and while one subunit binds GTP, the catalytic residue necessary to hydrolyze the terminal phosphate in GTP is provided by the adjoining subunit~\cite{Lowe1998, Wagstaff2017, Du2018}. GTP hydrolysis rates in FtsZ polymers regulate the turnover rates of the filaments~\cite{Anderson2004, Adams2011, Srinivasan2008} and, more recently, have been shown to drive treadmilling activity in FtsZ filaments~\cite{Ramirez2018, Yang2017, McCausland2021}. \textit{In vivo}, FtsZ filaments are dynamic. They are seen to organize as spiral structures along the axis of the cell during the cell elongation and condense into a ring (Z-ring) during cytokinesis, which is tightly coupled to the cell cycle progression~\cite{BENYEHUDA2002, Thanedar2004, Fu2010}. In \textit{Bacillus subtilis}, such spiral FtsZ filaments have been seen to coalesce into a ring during sporulation~\cite{BENYEHUDA2002}. Other studies have also suggested a spiral or helical intermediate in the formation of the cytokinetic FtsZ ring~\cite{Monahan2009}. 
}
However, despite numerous earlier studies, the exact mechanisms of FtsZ ring formation are not fully understood.

FtsZ is not only indispensable for cell division but is also the first protein to localize to the division site and orchestrates the recruitment of more than a dozen other proteins required for septation~\cite{Bi1991, Hale1997, Pichoff2002, Pichoff2005, Weiss1999}. FtsZ is also thought to be an ancient protein that originated as soon as cells formed during an early stage of genetic code evolution~\cite{Davis2002}. Thus, understanding FtsZ function and its assembly into the cytokinetic ring remains a fundamental question central to biology. FtsZ, therefore, has received extensive attention in the past decades, both in experiments~\cite{Stricker2002, Romberg2003, Anderson2004, Mahone2020, Walker2020} and in numerical studies~\cite{Ghosh2008, Reedy2000, Allard2008, Erickson2009a, Nguyen2021, MateosGil2019, Corbin2020}. Numerical studies in two dimensions showed that lateral attraction plays a crucial part in determining the final shape of the FtsZ filament~\cite{Horger2008}.
However, this model~\cite{Horger2008} lacked spontaneous torsion, which, as shown recently in Ref.~\cite{Ramirez2021}, is important in generating the required constriction force. 
In recent years, considerable advancements have been made in comprehending the role, function, and mechanisms of FtsZ, particularly how it produces constriction force on the membrane.

The FtsZ filaments constituting rings are found to be highly dynamic~\cite{Stricker2002, Anderson2004, Chen2005, Ramirez2018}. 
Recently, it has been shown that treadmilling not only guides the cell wall synthesis machinery~\cite{McCausland2021, Yang2017, Bisson-Filho2017, Whitley2024, Schäper2024} but also plays a crucial role in FtsZ ring assembly~\cite{Whitley2021}. 
FtsZ polymerizes and forms a large variety of dynamic structures, straight, bent, and helical filaments,  which are highly dependent on the experimental conditions such as pH,  the presence of various monovalent salts, crowding agents, and polycations~\cite{Popp2010}. 
Using the structure of crystallized \textit{Methanococcus jannaschii} FtsZ dimer and numerical simulations, the torsional and bending curvatures of FtsZ filaments were extracted~\cite{Pablo2014}.   
The intrinsic torsional curvature present in the FtsZ could deform the attached membranes via torsional stress and show chiral treadmilling vortices {\em in vivo}~\cite{Diego2018}. Earlier experiments also suggested that GTPase activity and concentration of free $\text{Mg}^{2+}$ ions play an important role in the dynamics and ring formation~\cite{Rivas2000, Monterroso2023}, and free $\text{Mg}^{2+}$ ions favor the lateral contacts between the filaments.

While these {\it in vivo} and {\it in vitro} reconstitution experiments have generated deep insights into the assembly dynamic of FtsZ filaments, a slightly different approach to studying bacterial cytoskeleton assembly in a molecularly crowded cellular environment was taken by Balasubamanian and colleagues~\cite{Srinivasan2008}. 
 The authors expressed FtsZ in yeast cells, both budding yeast (\textit{Saccharomyces cerevisiae}) as well as fission yeast (\textit{Schizosaccharomyces pombe}), and showed that FtsZ can assemble into a ring-like structure even in the absence of all other bacterial cytokinetic proteins~\cite{Srinivasan2008}. They further showed that the process of ring assembly involves a spooling-like mechanism, possibly driven by GTP hydrolysis and lateral associations among the FtsZ filaments.

{
In this paper, we aim to understand the mechanisms behind the emergence of observed FtsZ filament morphologies in both bacterial and yeast systems using a polymer model and Langevin dynamics simulations. Our theoretical approach assumes the existence of FtsZ filamentous assembly as the starting point. We model it as a tangentially active chiral semiflexible polymer characterized by finite bending and torsional rigidity, as well as intrinsic bending and torsional curvatures. A tangential activity is used to model the FtsZ treadmilling. \dc{We use a comparison with a simple treadmilling model to validate the effectiveness of our approach.}  
Our primary achievements are: 
$(i)$~Our model accurately captures all experimentally observed FtsZ morphologies, such as rings, helices, filaments, and globules, observed in FtsZ from different bacterial species or FtsZ mutants of {\em Escherichia coli}. 
$(ii)$~A minimal lateral attraction is necessary for ring formation, and treadmilling can flatten and widen globules into rings, displaying rotation and a spooling mechanism to ring formation. 
$(iii)$~The ring state remains stable over a broad range of treadmilling activities before transitioning into the open helix (OH) through phase coexistence. We obtain a detailed phase diagram, with the phase boundaries marked by peaks in size fluctuations and extensile response. 
$(iv)$~Active spinning chiral rings (SCR) induce contractile stresses, rotation, and shear that intensify near the phase boundaries.

 The paper is organized as follows: In Sec.~\ref{sec:experiment}, we first summarise the experimental results, setting the motivation for the theoretical study.  In Sec.~\ref{sec_model}, we present the model and simulation details. In Sec.~\ref{sec:results}, we discuss the main results, including the phase diagram, shape and size of the polymer, and stress generation.  Finally, we conclude in Sec.~\ref{sec:conclusion}, summarizing our results and presenting an outlook.

\section{Experimental results: motivation for theory} \label{sec:experiment}
{
The various shapes of FtsZ assemblies like rings,  helices, filaments, and condensates that have been found in several \textit{in vitro} experiments and \textit{in vivo} studies in bacteria have also been observed for several kinds of FtsZs expressed in fission yeast. Fission yeast has served as a useful model to study the assembly dynamics of ring formation by FtsZ ~\cite{Srinivasan2008}. It has also been useful to study and compare the response of FtsZ polymers from different species to drugs~\cite{Sharma2023}. Most importantly, the yeast system provides the ability to study the effect of mutations on FtsZ polymer morphology and dynamics, which would have been impossible otherwise due to their lethal effects on bacterial cells. FtsZ from different bacterial species or FtsZ mutants of \textit{Escherichia coli} have been shown to assemble into different forms such as helices~\cite{Sharma2023} (\textit{Helicobacter pylori} FtsZ or HpFtsZ), rings~\cite{Srinivasan2008, Sharma2023} (EcFtsZ, \textit{Staphylococcus aureus} SaFtsZ and EcFtsZ Q47K), globules (EcFtsZ D209A), open chains (EcFtsZ Q47K and EcFtsZ Q47KD86K), closed chains (EcFtsZ Q47KD86K) and network of polymers~\cite{Srinivasan2008} (EcFtsZ and EcFtsZ G105S). { In the experiments reported here, we have expressed FtsZ as C-terminal GFP fusions in fission yeast. We have utilized the medium strength thiamine repressible promoter containing vector, pREP42 ~\cite{Basi1993}. The construction of these plasmids has been already described~\cite{Srinivasan2008, Sharma2023}. To allow the expression and synthesis of FtsZ-GFP protein, we cultured the fission yeast cells carrying the various FtsZ-GFP variants in Edinburg Minimal Medium (EMM) in the absence of thiamine for 20 to 24 hours before imaging. We placed live yeast cells on agarose pads made with EMM and imaged them using a fluorescence microscope.} The different forms of FtsZ assemblies that have been observed upon expression in fission yeast are shown in Fig.\ref{fig_expt}($a$).
Further, since linear and helical polymers of FtsZ have been shown to undergo a transition into rings, or vice-versa in yeast and bacterial systems~\cite{Srinivasan2008, BENYEHUDA2002, Thanedar2004}, the forms observed for the FtsZ mutants in yeast cells might suggest that the mutations drive and stabilize one or the other states. Dynamics of EcFtsZ
Q47K ring formation in fission yeast ({\em S. pombe}) from an open chain shows a spooling mechanism. }

As revealed by independent electron microscopy studies~\cite{Popp2009}, the filaments observed in fluorescent microscopy Fig.~\ref{fig_expt}($a$) are typically filament bundles composed of multiple FtsZ chains bound laterally to each other. In the following section, we present a polymer model for such filament bundles with the intention of developing an understanding of the possible mechanisms controlling different emergent morphologies.

\begin{figure*}[t]
\centering
  \includegraphics[width=18cm]{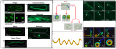}
\caption{{
($a$)~Images showing the various forms of FtsZ polymers when expressed in fission yeast. (i) ring-like structures of \textit{E. coli} FtsZ (EcFtsZ), (ii) helical or spiral polymers of \textit{Helicobacter pylori} FtsZ, (iii) straight linear polymers (open chains) of EcFtsZ Q47K D86K mutant, (iv) closed loops (closed chains) formed by EcFtsZ Q47K D86K mutant, (v) flexible polymers or networks formed by FtsZ G105S mutant and (vi) blobs or globules formed by the GTPase catalytic mutant EcFtsZ D209A. { Images in (i) and (ii) are maximal intensity projections of 3D-SIM (Structured Illumination Microscopy) images of $0.2 \mu m$ Z-planes acquired using a Deltavision OMX microscope. All other images are maximal intensity projections of $0.2 \mu m$ Z-planes acquired using a confocal laser scanning microscope.} The scale bar represents $3 \mu m$ { for all images, except the insets, where it is $0.5 \mu m$}.
($b$)~Schematic diagram of the active chains studied at different parameter regimes and lateral attraction (LA), recapitulating conformations in ($a$).
($c$)~Schematic representation of the active chain with active force $f_a$ acting tangentially along the chain's backbone. \dc{ {\em Spooling dynamics of ring formation:}   ($d$)~Dynamics of EcFtsZ Q47K ring formation in fission yeast (\textit{S. pombe}) from an open chain. Time-lapse imaging shows the different stages of ring assembly (time is in minutes). \dc{The scale bar represents $3\,\mu$m.} ($e$)~Configurations of the active chain in the presence of lateral attraction at $Pe=1$ and $\tilde \e = 1$ at different stages of ring formation with time $t_i$ captures the spooling dynamics in ($d$).}  In this study, all plots of the model polymer are generated using Ovito~\cite{ovito}, and the filament color coding from blue to red represents the filament's tail to head, the direction of treadmilling.
}}
\label{fig_expt}
\end{figure*}

\section{Model}
\label{sec_model} 

We consider an active bead-spring chain with bending and torsional rigidity to model the FtsZ filament. The polymer is denoted by a space curve described by position vectors $\{\rv_1, \rv_2\,\dots,\rv_N\}$ corresponding to $N$ beads.  The bonding energy is given by 
$U_b  = \frac{k_S}{2} \sum_{i=1}^{N-1} \left( |\bv_i| -r_0 \right)^2$ where the bond vector $\bv_i = \rv_{i+1} -\rv_i$, the rest-length of the bonds is $r_0$, and $k_S$ is the bond stiffness. %
The bending energy is given by
  $U_{\theta} = \frac{\kappa_{\theta}}{2}\sum_{i=1}^{N-2} \left(\theta_i - \langle\theta\rangle\right)^2$,
where $\h_i$  is the angle between two consecutive bond vectors $\bv_{i}$ and $\bv_{i+1}$, and $\k_\h$ denotes the bending modulus.
The non-planar nature of the space curve is described by the dihedral angle $\d_i$  between two successive planes constituted of bond-vectors $\{ \bv_{i},~  \bv_{i+1}\}$  and $\{ \bv_{i+1},~ \bv_{i+2} \}$ involving three successive segments.  
The torsional energy of the chain is modeled by  
  $U_{\phi} = \frac{\kappa_{\phi}}{2}\sum_{i=1}^{N-3} \left(\phi_i - \langle\phi\rangle\right)^2$,
where $\k_\phi$ is the torsional modulus. 

The self-avoidance and lateral attraction in the chain are modeled using an interaction potential $U_{NB}$ between non-bonded monomers, with
\begin{small}\begin{equation}
	U_{\textrm{\text{NB}}}(r) = 
  \begin{cases}
 \sum_{\la i,j\ra}
	4 \epsilon \left[ \left( \frac{\sigma}{r_{ij}} \right)^{12} - \left( \frac{
	\sigma}{r_{ij}} \right)^{6} \right] + \epsilon & \text{for} \ r_{ij} \le r_{\text{min}} \cr
			0  & \text{for} \ r_{ij} > r_{\text{min}}.
	\end{cases}
\end{equation}\end{small}
A choice of $r_{\text{min}}=2^{1/6} \sigma$ makes the above potential purely repulsive, the Weeks-Chandler-Andersen (WCA) interaction. This choice can model a self-avoidance. In addition, attraction between filament segments can be incorporated by replacing the distance cutoff by the typical Lennard-Jones (LJ) choice $r_{\text{min}}=2.5 \sigma$ so that monomers at a separation beyond the WCA cutoff attract each other via the LJ potential. The total energy cost is $ U =  U_{\textrm{b}} + 
U_{\theta} +  U_{\phi} + U_{\textrm{NB}}$.   

We model the treadmilling activity in terms of a tangential active force $f_a {\bf b}_i$ that acts along each bond. Here $f_a$ denotes a force per unit length~\cite{Holder2015}. This leads to a total self-propulsion force ${\bf F}_a = \sum_{i=1}^{N-1} f_a {\bf b}_i$. We use the dimensionless P{\'e}clet number,
\begin{equation}
Pe = \frac{f_a r_0^2}{k_BT},
\end{equation}
to characterize the local activity. Note that, for a chain of length $L=(N-1)r_0$, a related measure of activity $Pe_L=Pe (L/r_0)^2$ that scales quadratically with the chain length has been used in recent studies of active polymers~\cite{Holder2015}.

The Langevin dynamics of the chain are described by
			
			\begin{equation}
	m  \dot{\boldsymbol v}_i = - {\boldsymbol \nabla}_i U - \g
	{\boldsymbol v}_i + \g \sqrt{2D}\,{\boldsymbol \eta}_i + {\boldsymbol f}_{a}^i.
    \label{eq_lange}
\end{equation}
where, ${\boldsymbol f}_{a}^i = f_a {\bf u}_i$ with ${\bf u}_i= ({\bf b}_{i-1} + {\bf b}_i)/2$ for $i=2,\dots,(N-1)$ and ${\bf u}_1={\bf b}_1/2$, ${\bf u}_N={\bf b}_{N-1}/2$. In the absence of active drive, the above equation obeys equilibrium fluctuation-dissipation with monomer diffusivity $D=k_B T/\g$, and Gaussian random noise ${\boldsymbol \eta}_i$  satisfying $\la {\boldsymbol \eta}_i\ra=0$ and $\langle { \boldsymbol \eta}_i(t) \otimes {\boldsymbol\eta}_j(t^{\prime}) \rangle = \delta_{ij} \delta(t - t^{\prime}) \mathbb{1}$. 
The length, energy, and time scales are set by $\s=r_0$, $k_B T$, and $\t=\s \sqrt{m/k_B T}$, where $\t$ also describes the time scale for diffusion over $\s$. In these
units, we choose $k_S = 100$ and set $\kappa_{\theta}/\kappa_{\phi}=2.915$,  $\langle\theta\rangle = 7.6^{\circ}$ and $\langle\phi\rangle=20^{\circ}$ in accordance with previous all-atom simulations that were in agreement with properties of wild-type FtsZ filaments~\cite{Pablo2014}.  Further, we choose $\kappa_{\theta}=20$, such that the ratio of the chain length $L$ to the persistence length $\ell_p$ of the polymer, $L/\ell_p = 5$, consistent with measured persistence lengths~\cite{Srinivasan2008}~(also see Appendix-\ref{app_Llp}). The strength of the inter-segment attraction is tuned using the dimensionless control parameter $\tilde \e = \e/k_B T$. The molecular dynamics simulations are performed using $\d t = 10^{-4}\, \t$ within the velocity-Verlet implementation in LAMMPS~\cite{Plimpton1995} with in-house modifications to incorporate the tangential activity. \dc{The statistics are collected over $10^9$ time steps ($10^5\, \t$), once the system has reached a steady state, for subsequent analysis.  The parameter values used in the simulations are shown in Table-\ref{table_params}.}

\begin{table}[h!]
\dc{
\renewcommand{\arraystretch}{1}
\setlength{\tabcolsep}{5pt}
\small
\begin{tabular}{>{\raggedright\arraybackslash}p{5cm} p{2cm}}
\textbf{Parameters used} & \textbf{Values} \\
\hline
Number of monomers $N$ & $100$ \\ 
Temperature $T$ & 1.0 $k_B T$ \\
Time step $\Delta t$ & 0.0001 $\tau$ \\
Bond length $r_0$ & 1.0 $\sigma$ \\
Spring constant $k_S$ & 100 $k_B T / \sigma^2$ \\
Bending stiffness $\kappa_{\theta}$ & 20 $k_B T$~\cite{Srinivasan2008} \\
Intrinsic bending curvature $\langle \theta\rangle$ & 7.6$^\circ$~\cite{Pablo2014} \\
Torsional stiffness $\kappa_{\delta}$ & $\kappa_{\theta}/2.915$~\cite{Pablo2014}  \\
Intrinsic torsional curvature $\langle \phi\rangle$ & 20$^\circ$~\cite{Pablo2014} \\
Lennard-Jones potential depth $\epsilon$ & $0.1-3.0$ $k_B T$ \\
P{\'e}clet number& $10^{-6}-2$ \\
\end{tabular}
\caption{Parameters used for the polymer study in Langevin dynamics simulations, with length, energy, and time given in units of $\sigma$, $k_B T$, and $\tau = \sigma \sqrt{m / k_B T}$, respectively. The chain length $L=(N-1) r_0 \approx 10 \mu$m~(Appendix-\ref{app_Llp}).\label{table_params}} 
}

\end{table}

\section{Results}\label{sec:results}
FtsZ filaments exhibit a variety of structural forms, including helices, rings, open filaments, and globules, as shown in Fig.\ref{fig_expt}($a$) and captured within our model, Fig.\ref{fig_expt}($b$). In the absence of lateral attraction (WCA repulsion between segments) and activity, the model proposed above leads to a native state of equilibrium helical configurations~\cite{Pablo2014} \alert{(Supplemental Movie~1)}. This conformation can be seen in the top panel of Fig.\ref{fig_expt}($b$). 
Such morphologies can relax to simple equilibrium open chains in the absence of helicity in two limits of (i)~vanishing torsional rigidity or (ii)~intrinsic curvature $\langle\phi\rangle=\pi$ constraining polymeric fluctuations to an arbitrary plane \alert{(Supplemental Movies~2 and 3)}. Examples of such conformations are shown in the bottom right panels in Fig.\ref{fig_expt}($b$).  As mentioned before, Fig.\ref{fig_expt}($c$) depicts the tangential activity implemented in our model. 

On the other hand, large lateral attraction collapses the equilibrium filament into a compact globule. As mentioned before, the attraction between polymer segments is incorporated within our model by the full Lennard-Jones potential.  This is similar to the well-known coil-globule transition~\cite{deGennes}. However, the globule, in this case, comprises a compact arrangement of non-coplanar rings lying in different orientations \alert{(Supplemental Movie~4)}. 
In contrast, at an intermediate lateral attraction, the open helix first transitions into a ring \alert{(Supplemental Movie~5)}. In the presence of activity, the ring rotates in a chiral fashion to form spinning chiral rings~(SCR) \alert{: see  Supplemental Movies~6 and 7}. These dynamics distinguish them from the equilibrium rings and, as we show later, produce active contractile stress.  \dc{As depicted in Fig.~\ref{fig_expt}($e$) and \alert{Supplemental Movie 8}, the active helical chain at $Pe=1$ and $\tilde \e = 1$ transitions from an open helix conformation to SCR through a spooling-like mechanism as shown by EcFtsZ Q47K ring formation in heterologous yeast system in Fig.~\ref{fig_expt}($d$) \alert{(see Supplemental Movie~9)}.} This behavior was reported before by one of us in Ref.~\cite{Srinivasan2008}).

\begin{figure}[t!]
\centering
  \includegraphics[width=8cm]{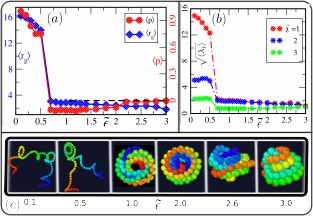}
 \caption{$(a)$ The dependence of the radius of gyration $\la r_g \ra$ and prolateness $\la p \ra$ on lateral attraction strength $\tilde \epsilon$ at equilibrium ($Pe=0$). The system exhibits an open conformation for low $\tilde \epsilon$, transitioning to ring and globule conformations for higher $\tilde \epsilon$. $(b)$ The behavior of the principal moments (eigenvalues) of the gyration tensor, $\lambda_i$'s, with respect to $\tilde \epsilon$. $(c)$ Representation of typical equilibrium configurations for different values of lateral attraction strength ($\tilde \epsilon$).  The system parameters are set as follows: bending modulus $\kappa_{\theta}=20$, intrinsic bending curvature $\langle\theta\rangle = 7.6^{\circ}$, and intrinsic torsional curvature $\langle\phi\rangle=20^{\circ}$. }\label{fig_RgVar}
\end{figure}

\subsection{Equilibrium helix to ring and globule}

With increasing inter-segment attraction $\tilde \e = \e/k_B T$, the open helix state of the polymer first undergoes a transition to a ring. At further higher attraction, the single ring further collapses into a globule consisting of multiple internal ring-like conformations of different orientations; see Fig.\ref{fig_RgVar}. The transitions are quantified utilizing the radius of gyration tensor~\cite{deGennes, Aronovitz1986, Cannon1991, Dima2004}
\bea
\mathcal{R}_{\a\be} = \f{1}{2 N^2} \sum_{i,j=1}^N (r_{i\a} - r_{j\a}) (r_{i\be} - r_{j\be}) 
\eea
for a chain of $N$ beads with  $r_{i\a}$ denoting the $\a$-th component of the position vector ${\rv}_i$ of $i$-th bead. With $\l_1 \geq \l_2 \geq \l_3$ denoting the three eigenvalues of $R_{\a\be}$, the principal radii of gyration, 
and using 
\bea \label{eqn:Rg}
&\l_{tr}= \l_1+\l_2+\l_3,
\eea
 we can estimate the mean polymer size as { $\la r_g \ra= \sqrt{\la \l_{tr}\ra}$,} and use the prolateness 
\bea \label{eqn:S}
p = \f{\prod_{i=1}^3 (3 \l_i - \l_{tr})}{\l^3_{tr}},
\eea
to obtain {the mean prolateness $\la p \ra$ averaging over all steady-state configurations.}
A related positive quantity, asphericity, which vanishes for a spherically symmetric shape, is utilized later in this paper. 
 The mean prolateness $\la p \ra$ is bounded above and below: $-1/4\,\leq \la p \ra \leq 2$, with positive (negative) values of $p$ denoting prolate (oblate) shapes, and  $\la p \ra=0$ for isotropic conformations. 

 The polymer size shows a sharp drop across the transition of an open helix to an equilibrium ring at $\tilde \e = \tilde \e^\ast$; see Fig.\ref{fig_RgVar}($a$). At higher $\tilde \e$, the size $r_g$ keeps on decreasing, although ever so little. This decrease in size is associated with another structural `transition', from ring to compact globule. This transition is further quantified in terms of the prolateness $\la p \ra$, which grows from negative values~($\la p\ra=-0.1$ at $\tilde \e^\ast$) for the ring to vanish for the spherically symmetric globule. Note that, here, the ring state cannot be a two-dimensional object but must have a height due to the width of the filament and self-avoidance. Thus, the ring state has approximately the shape of a short pillbox with diameter $\l_1=\l_2 \equiv \l$ and height $\l_<$. Within this approximation, one gets 
 \bea
 \la p \ra=-\f{1}{4} \left\la 
\left( \f{1 -\f{\l_<}{\l} } {1 + \f{\l_<}{2\l} }\right)^3\right\ra
 \approx -\f{1}{4} \left(1 -\left\la\f{9\l_<}{2\l} \right\ra \right), 
 \label{eq_S}
 \eea
 linearizing for small parameter $\la \l_</\l \ra$. Thus, in the limit of vanishing height with respect to the ring size, $\la \l_</\l \ra \to 0$, one expects $p \approx -0.25$.  The numerical simulation results in Fig.\ref{fig_RgVar}($a$) show the actual $p$ value at the transition point remains above this estimate, suggesting a non-zero $\la \l_</\l \ra \approx 2/15$, an estimate that agrees approximately with $\la \l_3/\l_1 \ra$ at $\tilde \e^\ast$ shown in Fig.\ref{fig_RgVar}($b$). With increasing $\tilde \e$ beyond this transition point, $\la p \ra$ increases as the ring transitions towards a globule to vanish at $\tilde \e=3$, identifying the spherical symmetry of the final globule state.

\begin{figure*}[t!]
\centering
  \includegraphics[width=16cm]{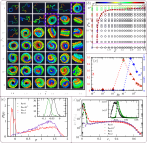}
\caption{$(a)$~Configurational phase diagram in the $Pe-\tilde \epsilon$ plane. The active chains are color-coded from the tail (blue) to the head (red), with self-propulsion directed from the tail to the head.  
Note that in terms of an alternative measure $Pe \,(L/r_0)^2$, the polymer activity in the figure ranges from 
$0$ to $2\times 10^4$.
$(b)$~Quantitative phase diagram in the $Pe-\epsilon$ plane using fluctuations $\sigma^2 = \langle \tilde r_{e}^2\rangle-\langle \tilde r_{e}\rangle^2$ of the scaled end-to-end separation $\tilde {r}_e = r_e/L$ represented by points denoted by colors as shown in the color bar. 
$(c)$~Plot of $\sigma^2$ as a function of $Pe$ for a fixed $\tilde \e=1$ and shows a maximum at $Pe^*=0.5$ and decreases away from this point. The plot is shown by blue ($\diamond$) data points and the axis label on the right. The extension modulus $\chi$ is plotted with the red-colored left axis and red ($\triangle$) data points. 
$\chi$ exhibits a non-monotonic increase with activity $Pe$ and peaks near the transition point $Pe^{**}=0.1$.
The green-colored symbol denotes a representative error bar. 
$(d)$~Probability distributions $p(\tilde r_e)$ 
for various $Pe$ values at $\tilde \e =2$. The clear bimodality in the distribution function at $Pe=1$ shows the coexistence of compact globule and ring structures with open helical conformation. Inset:~compares the distributions at $Pe=0$ and $10^{-3}$.
$(e)$~Probability distributions $P(p)$ of the prolateness $p$ for the same $Pe$ values as in $(d)$ and $\tilde \e =2$. At $Pe=1$, the distribution shows maxima corresponding to globule (peak around $p=0$), ring (peaks at $p \approx -0.1$ and $-0.15$), and open helix (peak for $p\approx 1.5$).
}
\label{fig:ResultMain}
\end{figure*}

 These transitions can be directly read off from the variation of $\l_i$s in Fig.\ref{fig_RgVar}($b$). In the open helix state, all the eigenvalues are different from each other. At the formation of the spiral, two larger eigenvalues corresponding to the diameter of the ring become equal, $\l_1=\l_2$, while the height of this pillbox-shape $\l_3$ remains smaller. With increasing $\tilde \e$, $\l_1=\l_2$ shrinks, and the $\l_3$ increases to become equal at the formation of spherical globule at $\tilde \e=3$. Note that the transformation of the ring state to the globule is gradual and does not show any signature of phase transition.  As we show in the later section, the presence of treadmilling activity can transform a globule into a ring, thereby potentially stabilizing rings in the biological context. 

\subsection{Role of treadmilling activity}
The shape of the equilibrium ring formed at $\tilde \e \approx 1$ remains intact over a wide range of treadmilling activity $Pe$. Before opening up, the ring shows chiral rotation, characterizing the non-equilibrium  SCR state. However, for higher $\tilde \e$ that stabilize globular structures at equilibrium, activity helps to unwind the globule to form rotating rings, thereby stabilizing SCR. 
At even higher $Pe$, the rings do unfold into open helices. For example, at $\tilde \e \approx 1$, the ring opens up into a helix above $Pe \approx 0.1$. 

We present a comprehensive phase diagram in Fig.~\ref{fig:ResultMain}$(a)$ and ($b$) as a function of treadmilling activity $Pe$ and lateral attraction strength $\tilde \epsilon$. Fig.~\ref{fig:ResultMain}$(a)$ illustrates representative conformations of the chain at different parameter values. This plot shows representative conformations at $\tilde \e \geq 0.8$ where the equilibrium chain is already in a ring or globule state. For a similar but more detailed phase diagram, see Appendix-~\ref{appendA}.  At any given $\tilde \epsilon$, the ring destabilizes at high enough $Pe$ to form an open helix. At small to intermediate $\tilde \epsilon$, the figure suggests that the ring state remains stable over a broad range of $Pe$. However, at high enough $\tilde \epsilon$ (e.g., at $\tilde \epsilon\approx 2.4$), the equilibrium chain is in a globule state. With increasing $Pe$, the globule first transitions to SCR before undergoing a transition to open helix at high enough $Pe$~($\approx 1$) as shown in Fig.~\ref{fig_PeS}

 \begin{figure}[t!]
\centering
  \includegraphics[width=8cm]{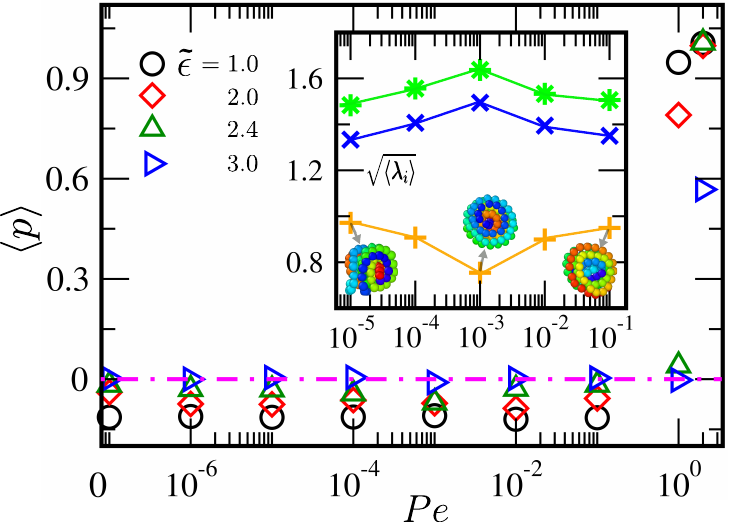}
\caption{The mean prolateness $\la p \ra$ as a function of activity $Pe$ for various $\tilde{\epsilon}$ values. The globule formed at equilibrium condition ($Pe=0$) transitions into a ring shape as a function of activity ($Pe$). Inset: Variation of principal radii of gyration, $\lambda_i$'s, with $Pe$ at $\tilde \e=2.4$, with symbols for $i=1$~(\textcolor{green}{$\ast$}), $i=2$~(\textcolor{blue}{$\times$}) and $i=3$~(\textcolor{MyOrange}{$+$}). 
}
\label{fig_PeS}
\end{figure}

In Fig.~\ref{fig:ResultMain} (b), we present a quantitative version of the phase diagram using fluctuations in the chain's end-to-end separation. The color bar indicates the amount of  fluctuations $\s^2 = \la \tilde r_e^2\ra - \la \tilde r_e\ra^2$ in the scaled end-to-end separation $\tilde r_e = r_e/L$,
where $L$ is the chain length. 
Near the transition from ring to open helix denoted by the dashed red line in Fig.~\ref{fig:ResultMain}(b), $\sigma^2$ reaches a maximum.   Fig.~\ref{fig:ResultMain}(c) shows the variation of $\sigma^2$ as a function of $Pe$ for a fixed $\epsilon=1$ (blue colored points and ordinate label) with the maximum at $Pe^*=0.5$, the transition point. It sharply diminishes away from this transition point.

This figure also shows the linear response $\chi$ of the polymer extension under external force~(see Appendix-\ref{appendB} and Fig.\ref{fig:ResFunc}).
It varies non-monotonically with $Pe$ and shows a maximum that precedes the maximum in $\s^2$, e.g., at $Pe^{**}=0.1$ for $\e=1$. 
The maximum is attributable to the instability of the ring structure near the transition, where even a small amount of extensile force is sufficient to destabilize the ring and open it.

In Fig.~\ref{fig:ResultMain}$(d)$ and $(e)$, we present the probability distributions of end-to-end separation $P(\tilde r_e)$ 
and prolateness $P(p)$ respectively, at various $Pe$ keeping $\tilde{\epsilon}=2$ fixed. The distribution functions are normalized to  $\int_0^1 d\tilde{r}_e\, 4 \pi \tilde{r}_e^2 \,P(\tilde{r}_e)=1$ and $\int dp\, P(p)=1$. At $Pe=0$ and $10^{-3}$, where the ring phase is stable, the distribution shows a single peak near $\tilde{r}_e \approx 0.03$; see Fig.~\ref{fig:ResultMain}$(d)$ and its inset. The compact structure at equilibrium swells under activity $Pe=10^{-3}$ to become more ring-like.
Near the transition point, $Pe=1$, the chain shows bimodality in $P(\tilde r_e)$, indicating the coexistence of open helix (maximum near $\tilde {r}_e = 0.45$) and compact ring (maximum near $\tilde {r}_e \lesssim 0.01$).
At higher $Pe$, the distribution becomes unimodal with a maximum around  $\tilde {r}_e = 0.45$ for $Pe=2$, corresponding to the open helix. The broad distribution characterizes large deviations around this maximum, allowing the chain to visit other conformations.

The distribution of prolateness $p$ remains unimodal at $Pe=0$ and $10^{-3}$ (Fig.~\ref{fig:ResultMain}$(e)$ and inset). The location of the maximum shifts towards smaller $p$ values at $Pe=10^{-3}$, with respect to equilibrium. This is due to the reorganization to a better ring via the non-equilibrium swelling. At $Pe=1$, the multimodal distribution function $P(p)$ displays the coexistence of ring, globule, and open helix, captured by the peaks at $p<0$, $p\approx 0$, and $p> 0$, respectively. Moreover, at this P{\'e}clet, the distribution shows a second maximum with split peaks at $p< 0$. The first one near $p=-0.1$ corresponds to the stable ring phase. The second one at $p < -0.1$ corresponds to a more swollen ring that is formed on the way of opening up from the ring state to the open helical conformation. The reduction of $p$ with swelling can be understood using larger $\l$ in Eq.\eqref{eq_S}.  
At a higher activity, $Pe=2$, as the chain opens up, the distribution becomes unimodal again with one maximum near $p\approx 1.25$. \dc{The multimodality in $P(\tilde r_e)$ and $P(p)$ signifies phase coexistence, a characteristic of first-order transition.}

\subsubsection{\dc{Analyzing active stabilization of ring using prolateness}}
The variation of mean prolateness $\la p \ra$ with $Pe$ corresponding to different $\tilde \e$ is shown in Fig.\ref{fig_PeS}. The equilibrium ring state at $\tilde \e=1$ remains relatively unaltered up to $Pe \approx 0.1$, where the ring transitions into an open helix. On the other hand, $\la p \ra$ increases with $\tilde \e$ at $Pe=0$ to vanish at large $\tilde \e$, as the equilibrium ring gets into globule-like shape; see Fig.\ref{fig_RgVar} as well. With increasing $Pe$, the equilibrium globules at $\tilde \e=2$ and $2.4$ become ring-like, showing a decrease in $\la p \ra$ reaching a negative minimum at an intermediate value of $Pe$ ($Pe=10^{-3}$) before increasing with activity as the rings open up. 
This suggests that treadmilling activity can regulate the formation of FtsZ rings. This feature is further characterized using the three principal radii of gyration $\la \l_i \ra^{1/2}$ with $i=1,2,3$, for $\tilde\e=2.4$ shown in the inset of Fig.\ref{fig_PeS}. At $Pe=10^{-3}$, $\la \l_1 \ra$ and $\la \l_2 \ra$ becomes largest as the ring widens and associated with this $\la \l_3 \ra$ gets to the smallest value signifying the maximal thinning of the ring. 
\dc{An equilibrium globule, stabilized by lateral attraction, can be flattened by the combined effects of treadmilling activity and intrinsic asphericity. Together, these factors can break the spherical symmetry, transforming the shape into a ring.}

\dc{
\subsubsection{Morphological transition and the size and shape of the polymer chain}\label{sec:size}
\begin{figure}[!hbt]
\centering
 {\includegraphics[width=8cm]{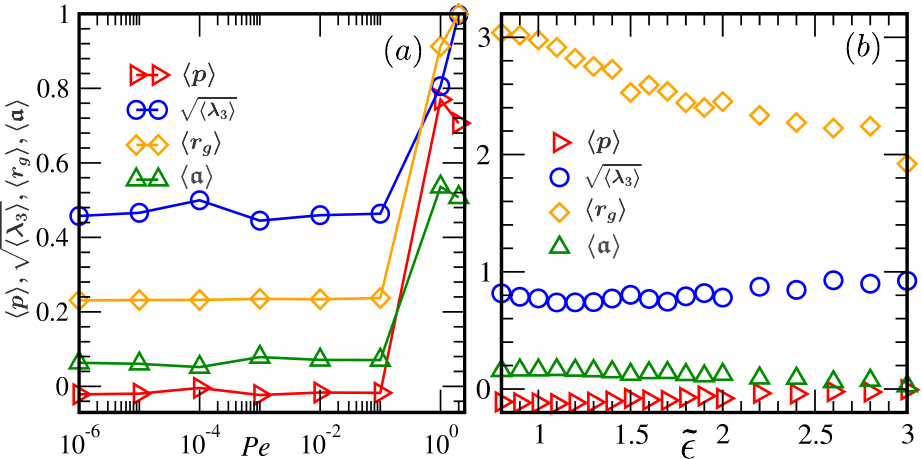}}
\caption{\dc{The mean radius of gyration $\la r_g \ra$, asphericity $\la \mathfrak{a} \ra$, thickness $w=\sqrt{\la \lambda_3 \ra}$ where $\l_3$ is the smallest eigenvalue of the radius of gyration tensor and prolateness $\la p \ra$ as function of $Pe$ for fixed lateral attraction $\tilde \epsilon=1$ ($a$) and as a function of $\tilde \epsilon$ for fixed $Pe=10^{-2}$ ($b$). In panel ($a$), the radius of gyration is scaled by the maximum $r_g^{max}=14.39$ and thickness by the maximum 
$w^{max}=2.4$ that are the maximum values of $r_g$ and $w$ 
corresponding to $Pe=2$. }
}\label{fig:shape1}
\end{figure}

The change in mean size and shape of the active chiral chain, \dc{and the morphological transition,} can further be analyzed in terms of the mean radius of gyration, asphericity, and prolateness. 
As we mentioned before, the size and shape of a polymer can be quantified using the eigenvalues of the radius of gyration tensor, specifically using the radius of gyration $r_g$ (Eq.~\ref{eqn:Rg}) and prolateness $p$ (Eq.~\ref{eqn:S}). In addition, here we use the asphericity $ \mathfrak{a}$ of a configuration which is given by 
\bea
 \mathfrak{a} = \f{3}{2}  \f{\sum_{i=1}^3 (\l_i - \f{\l_{tr}}{3})^2}{\l^2_{tr}} 
= 1 - 3  \f{\l_d}{\l^2_{tr}} 
\eea
where $\l_d = \l_1 \l_2 + \l_2 \l_3 + \l_3 \l_1$. The mean asphericity $\la  \mathfrak{a}\ra$ is obtained by averaging over steady-state configurations. For a perfect isotropy $\la  \mathfrak{a}\ra=0$ and $\la p \ra=0$, and the asphericity $ \mathfrak{a}$ is bounded above and below $0\leq  \mathfrak{a} \leq 1$.

In Fig.~\ref{fig:shape1}$(a)$, we show the variation of all these parameters as activity increases at  $\tilde\e=1$. For a wide range of activities ( $10^{-6} \leq Pe \leq 10^{-1}$) rings are stable with $\la r_g \ra$, $\la p \ra < 0$, $ \la \mathfrak{a} \ra >0$ remaining unchanged. Note that $\la \l_3\ra$ is a measure of ring thickness in the SCR phase. This also remains unchanged up to $Pe=0.1$. At high enough activity, $Pe > 0.1$, intersegment attraction gets overpowered by activity to break contacts between the monomers and destabilize the ring to an open helix state. 
Examples of the equilibrium ring and active rotating SCR are shown in \alert{Supplemental Movie 6,7}. In Fig.~\ref{fig:shape1}$(b)$, we show the variation of the same size and shape parameters as a function of $\tilde\e$ at a fixed $Pe=10^{-2}$. Increasing $\tilde\e$ shrinks the size $\la r_g \ra$. A clear crossover from ring to globule is captured by the vanishing of $\la p \ra$ and $ \la \mathfrak{a} \ra$.  
}

\subsubsection{\dc{Robustness of turning number}}
The number of turns in the SCR does not change before it opens up; however, the robustness of the ring state varies non-monotonically. This is established using the tangent-tangent correlation function $C_{\bt \bt}(s)= \la \bt (s) \cdot \bt (0)$ along the chain contour, and its Fourier transforms $S_{\bt \bt}(q)$; see Fig.\ref{fig_stq}. The correlation function shows periodic variation with the periodicity captured by the pronounced maximum in $S_{\bt \bt}(q)$ with the location of the maximum near $q=n/L$ denoting the number of turns $n$. Fig.\ref{fig_stq} shows the maximum is at $n=qL\approx 6$. As we see, this position of maximum does not change with $Pe$; however, the height of $S_{\bt \bt}(q)$ at the maximum varies non-monotonically. The initial increase with $Pe$ characterizes the formation of a more stable ring. The subsequent decrease of this peak height at even larger $Pe$ signifies the destabilization of the ring.

\subsubsection{Dynamics near the SCR-open helix phase boundary}
\begin{figure}[t!]
\centering
  {\includegraphics[width=7cm]{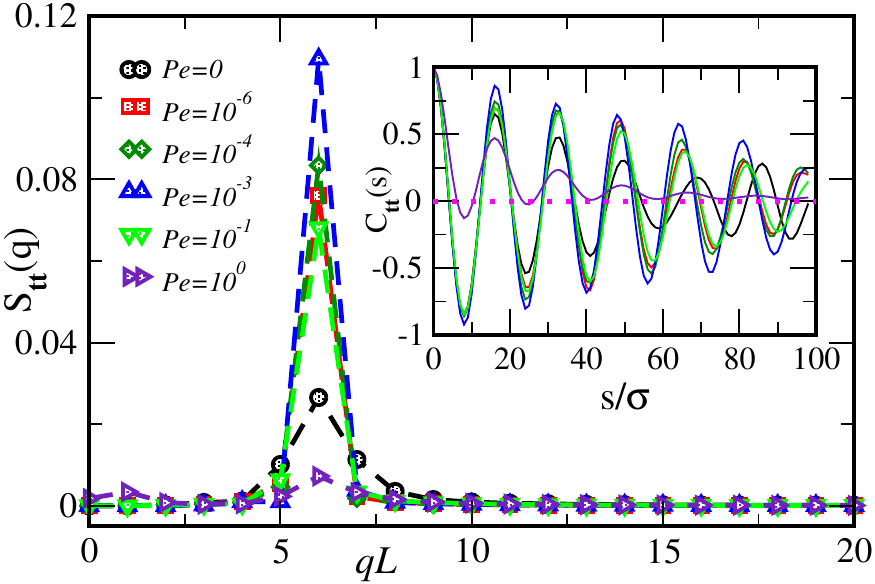}}
\caption{Fourier transform $S_{\bt \bt}(q)$  of the tangent-tangent correlation function $C_{\bt \bt}(s)$ for the helical chain for various $Pe$ values at fixed lateral attraction $\tilde \epsilon=1.0$. The correlation function is oscillatory for the ring phase, with the periodicity captured
by the peak in the structure factor. The peak height quantifies the degree of helicity of the chain, while the width of the peak indicates the statistical dispersion of the structure.}\label{fig_stq}
\end{figure}
The dynamics of the chain near the spiral to open helix transition can be analyzed using the time series of the principal radii of gyration $\l_i$ with $i=1,2,3$ and the prolateness $p$; see Fig.~\ref{fig_Dynamics}. The time series in Fig.~\ref{fig_Dynamics}($a$) \dc{considers a short time window between $t=1000 \t$ to $3000 \t$ from the full time series over $10^5 \t$ to highlight the intermittent dynamics around phase transition in which the system dynamically crosses over between different morphologies including} compact ring (denoted by $R1$), swelled ring ($R2$), globule ($G$), and open helix ($OH$) \alert{: see Supplemental Movie~10}. 
The globule is characterized by $\l_1=\l_2=\l_3$ and $p=0$. The data at the time point denoted by $G$ shows approximate agreement with this expectation. In the ring state, $\l_1 \approx \l_2$ is significantly larger than $\l_3$ and $p(R2) < p(R1)$ with both R1 and R2 rings having negative $p$. For OH, $\l_1 \gg \l_2(\approx \l_3)$ and $p \to 2$ (also see Fig.\ref{fig_CorrS} in Appendix-\ref{append_tseries}). 

 \begin{figure}[t!]
\centering
   \includegraphics[width=8cm]{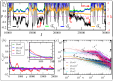}
\caption{$(a)$~A representative time series of $p$ and the principal radii of gyration at $Pe=1$ and $\tilde \epsilon =2$ where ring (marked with $R1$, $R2$), globule ($G$) and open helix ($OH$) co-exist. 
Here $R1$ and $R2$ represent the compact and swollen ring states with $p=-0.1$ and $p=-0.15$, respectively. 
$(b)$~Two-time auto-correlation function of $p$ for various $Pe$ values at $\tilde \epsilon =2$.
$(c)$~Power spectral densities $S_{pp}(\nu)$ for the same $\tilde \e$ and $Pe$ values as in $(b)$. The scatter plots denote the direct Fourier transform, and the solid lines are guides to the eye. The black dashed, dash-dotted and dotted lines represent $\nu^{-1}$, $\nu^{-2}$ and $\nu^{-3}$, respectively.
}
\label{fig_Dynamics}
\end{figure}

These dynamics can be further characterized in terms of two-time correlations and the corresponding power spectral density.  \dc{Note that the calculation of correlation function and power spectral density at a given parameter value uses the full steady-state time series over $10^5 \t$ and represents the system's dynamical property corresponding to the control parameters, e.g., $Pe$ and $\tilde \e$.} Fig.~\ref{fig_Dynamics}($b$) shows the correlation function $C_{pp}(t)= \la \d p(t) \d p(0)\ra/\la \d p^2(0)\ra$ with $\d p(t)=p(t)-\la p\ra$. The corresponding power spectrum $S_{pp}(\nu)$ is obtained by Fourier transforming the correlation function and is plotted in Fig.~\ref{fig_Dynamics}($c$). We used TISEAN~\cite{tisean} to compute the correlation and power spectrum. The apparent oscillations in $C_{pp}(t)$ at  $Pe=1$ correspond to the phase coexistence. The absence of any characteristic maximum in frequency $\nu$ in the power spectrum $S_{pp}(\nu)$ characterizes the absence of any robust oscillation in the dynamics.    
The high frequency behavior of $S_{pp}$ shows approximate power-law decay as $\nu^{-\beta}$ with $\beta=1$ in the stable SCR phase ($Pe=0$ and $10^{-3}$), $\be=2$ for  $Pe=1$ near the transition, and $\be=3$ for $Pe=2$ corresponding to open helix.

 \begin{figure}[t]
\centering
    \includegraphics[width=8cm]{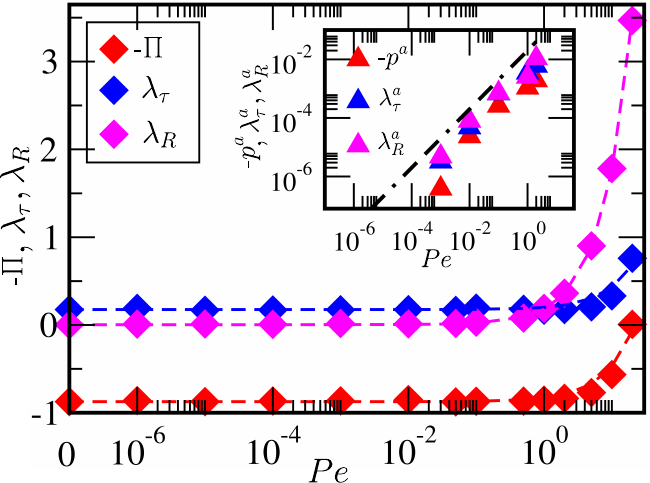}
\caption{
\dc{{Main plot:} Stress components as a function of $Pe$ in semi-log scale. The excess pressure $\varPi$,  magnitude of the largest eigenvalue of shear $ \l_\t$, and the largest eigenvalue of rotation $ \l_R$ increase with $Pe$ linearly from their equilibrium values $-\varPi^{eq}=0.873$, $ \l_\t ^{eq} = 0.176$ and $\l_R^{eq}=0.003$.  The dashed lines through data represent linear fits to the form $f(Pe) = f^{eq} + A_f Pe$, with $f^{eq}$ denoting equilibrium values and the fitted slope for pressure $A_{\varPi} = 0.04$, for shear $A_{\lambda_\tau} = 0.025$, and for rotation $A_{\lambda_R} = 0.174$. {\em Inset:} Plots pressure, shear, and rotation obtained from swim stress in Eq.\eqref{eq:swim} as a function of $Pe$ in log-log scale. The dash-dotted line indicates linear growth as $\sim Pe$.}}
\label{fig:StressGrid}
\end{figure}

{
\subsubsection{Stress generation by the spinning chiral ring}
\label{sec:stress}
Here, we present a direct measure of the active stress due to the spinning chiral ring (SCR). Recent advances led to derivations of swim pressure for active Brownian particles~\cite{Takatori2014, Yang2014,  Julian2015, Solon2015PRL}. Instead, we employ an embedding environment \dc{as a probe} to calculate the stress due to the active chain.  

For this purpose, we consider a harmonic cubic lattice of size $N_g^3$ with fixed boundaries to embed the SCR. The lattice spacing is chosen to be $\s$ and $N_g=17$ such that the box size $N_g \s$ is significantly larger than the diameter of the SCR so that the distortions do not reach the boundaries. The monomers of the chain are assumed to be bonded with lattice points within the cutoff distance $\s$. To avoid introducing new parameters, we choose these bond strengths and that of the harmonic lattice to be $k_S$, the same as in the active polymer. 

We calculate the excess stress on the cubic lattice due to the active chiral chain
\bea
\S_{\a\be} = - \f{1}{{\cal N} \s^3} \sum_{i=1}^{\cal N} \langle r_{i\alpha} f_{i\beta}\rangle\, ,
\eea
where ${\cal N}$ denotes the number of lattice points bonded with the chain, and ${\cal N} \s^3$ is the corresponding volume occupied by them. Here, $ f_{i \be}$ stands for the $\be$-th component of the force acting on the lattice point $i$ due to the chain.

The stress tensor $\Sigma_{\alpha \beta}$ is fully characterized by its three irreducible parts, the excess pressure $\varPi$, shear $\tau_{\alpha \beta}$, and rotation $R_{\alpha \beta}$. They are expressed as
\bea
    \varPi &=& -\frac{1}{d} \Sigma_{\alpha \alpha} \nn\\
    \tau_{\alpha \beta} &=& \frac{1}{2} (\Sigma_{\alpha \beta}+\Sigma_{\beta \alpha}) - \frac{1}{d} \Sigma_{\alpha \alpha} \delta_{\alpha \beta} \nn\\
    R_{\alpha \beta} &=& \frac{1}{2} (\Sigma_{\alpha \beta}-\Sigma_{ \beta \alpha}),
    \label{eq_irreducible}
\eea
where $d$ is the embedding dimension. The Einstein summation convention is assumed above. The pressure is a scalar, $\t_{\alpha \beta}$ is a traceless symmetric matrix, and $R_{\alpha \beta}$ is a purely antisymmetric matrix. 
In our study $d=3$. Thus, in 3d, shear $\t_{\alpha \beta}$ has five independent components, and rotation $R_{\alpha \beta}$ has three.
 
In Fig.~\ref{fig:StressGrid}, we show the dependence of the  $\varPi$, the magnitude of the largest eigenvalues $\l_\t$ of $\t_{\alpha \beta}$ and the largest eigenvalue $\l_R$ of $R_{\alpha \beta}$ as a function of activity. 
\dc{All of these quantities grow linearly with $Pe$; they show good fits to 
$-\varPi = -\varPi^{eq}+  A_{\varPi}\, Pe$, 
$\l_\t = \l_\t^{eq} + A_{\lambda_\tau} Pe$,
and
$\l_R = \l_R^{eq} + A_{\lambda_R} Pe$ where $\varPi ^{eq}$, $\l_\t^{eq}$ and $\l_R^{eq}$ denote the equilibrium values; see Fig.\ref{fig:StressGrid}. The deviation from equilibrium pressure, $\varPi-\varPi^{eq}=-A_{\varPi}\, Pe$ with $A_{\varPi}=0.04 >0$, indicates generation of an active contractile stress by SCR. The linear increase of all stress components with $Pe$ can be understood by considering the dimensionless swim stress from active Brownian motion~\cite{Takatori2014, Yang2014, Julian2015, Solon2015PRL}, 
\bea 
\sigma^{sw}_{\a, \be} = \frac{Pe}{V} \sum_{i=1}^N \la  u_{i, \a} r_{i, \be} \ra
\label{eq:swim}
\eea
where $V= r_g^3$ can be a natural choice of polymer volume. All lengths are expressed in units of bond-length $r_0=\s$.  Unlike in active Brownian particles, the persistence in motion of individual bonds in the polymer is an emergent property controlled by its mechanics. The above expression can be utilized to obtain swim pressure, shear, and rotation as in Eq.\eqref{eq_irreducible}. In the inset of Fig.\ref{fig:StressGrid}, we show the plots of swim pressure $p^{(a)}$, and the largest eigenvalues corresponding to shear $\l_\t^a$ and rotation $\l_R^a$. Note that the swim stress grows linearly with $Pe$ capturing the property of measured active stress in the main plot of Fig.~\ref{fig:StressGrid}.  
}

\dc{
\section{Discussion}
In our study so far, we have considered a model of polymer with fixed length with tangential activity. Here we present a critical analysis of the model asking two questions: (i)~how relevant is the model considered to a treadmilling chain, (ii)~what is the chain length dependence of the results obtained in terms of the different morphological phases and transitions. 

  \begin{figure}[t!] 
\centering
  \includegraphics[width=8 cm]{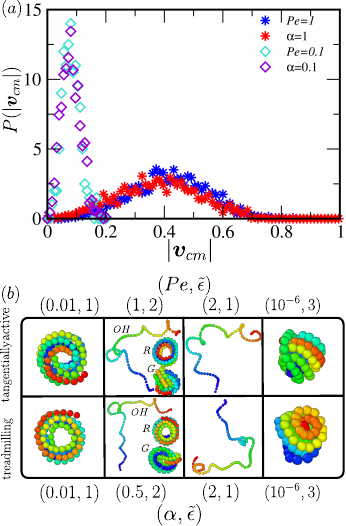}
\caption{ \dc{
$(a)$ 3D: Comparison of the steady-state distributions of the center-of-mass speed $|{\bf v}_{cm}|$ for the active polymer and treadmilling chain model without lateral attraction. Data presented at two sets of parameter values ($\a,\, Pe$) are denoted in the figure.  
$(b)$~Representative steady-state configurations for four distinct $(Pe, \tilde{\epsilon})$ and $(\alpha, \tilde{\epsilon})$ pairs for the active chain and treadmilling filament models. From left to right, figures show SCR, coexistence, open helix, and globule structures.
}
}
\label{fig:ShapeN100TmtAm}
\end{figure}

 \begin{figure}[t!] 
\centering
  \includegraphics[width=7 cm]{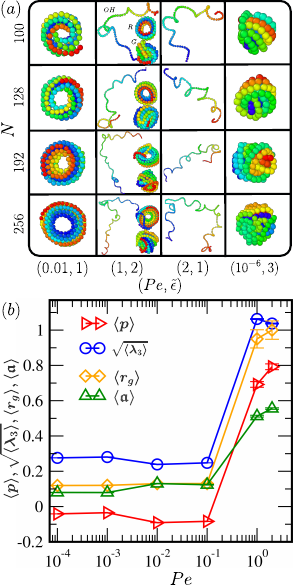}
\caption{\dc{
$(a)$ Representative steady-state configurations for four distinct $(Pe, \tilde{\epsilon})$ combinations at different chain lengths $N$, showing that the emergence of the ring, globule, open helix, and coexisting morphologies are determined by the parameter values and not the chain length. 
$(b)$~For a chain of $N=256$, the mean radius of gyration $\la r_g \ra$, asphericity $\la \mathfrak{a} \ra$, thickness $w=\sqrt{\la \lambda_3 \ra}$ and prolateness $\la p \ra$ as function of $Pe$ obtained using $\tilde \epsilon=1$. The radius of gyration and thickness are normalized by their respective maximum values, $r_g^{max}=28.3$ and $w^{max}=4.71$, obtained at $Pe=2$.
}
}
\label{fig:ShapeN256}
\end{figure}

\subsection{Comparison with treadmilling model}
In a treadmilling polymer, the chain grows at one end and shrinks at the other through stochastic addition and removal of monomers~\cite{Phillips2012, Erlenkamper2013}. The rates of addition $r_+$ and removal $r_-$ control the polymer length and its motion. Indeed, the observed chain length of FtsZ bundles in our fission yeast experiments showed values between $8$ to $13\, \mu$m with typical lengths remaining near $10\, \mu$m.   In the limit of $r_+ = r_- = \a$, the length distribution gets sharply peaked with a small variance. For a precise length control, let us consider a treadmilling filament with a turnover rate $\a$ in which a monomer of size $r_0$ from the tail end is removed and inserted at the head. This keeps the length unchanged and the motion of the filament in one dimension has the mean velocity $\bar v = \a r_0$. This active motion can be interpreted in terms of tangential activity $Pe= \bar v r_0/D$ for the active polymer model we considered. A comparison between the velocity distributions of the two models is shown in Fig.~\ref{fig:ShapeN100Tm}$(a)$ in Appendix-\ref{appendTreadmil}. In three dimensions (3D), in addition to the turnover rate, the velocity of the polymer center of mass ${\bf v}_{cm}$ is governed by the dynamic changes in polymer conformations. To compare the two models in 3D, we use chains in the absence of lateral attraction (WCA repulsion between non-bonded segments). We find a close agreement between the speed distributions from both models as shown at two sets of parameter values $\a$ and $Pe$ in  Fig.~\ref{fig:ShapeN100TmtAm}$(a)$.

In Fig.~\ref{fig:ShapeN100TmtAm}$(b)$, we compare typical steady-state configurations obtained from the tangentially active model and the treadmilling model. For comparable parameter values corresponding to SCR, globule, and open helical structures for active chains, the treadmilling polymer model yields identical results. It shows that the two models produce equivalent phenomenology with similar parameter values identifying the transition. However, the transition point characterized by intermittent dynamical crossovers between open chain, globule and SCR appears at a slightly smaller turnover rate $\a=0.5$ for the treadmilling model, with respect to that in the active chain $Pe=1$ in the presence of lateral attraction $\tilde \e=2$. At $\a=1$ corresponding to $Pe=1$, the treadmilling filament gets into the open helix state with transient local rings running from end to end~(see Fig.\ref{fig:ShapeN100Tm}$(b)$ in Appendix-\ref{appendTreadmil}).

\subsection{Chain length dependence of morphologies and transitions}
\label{appendSystemSize}
The results we presented so far for tangentially active chains were for a single length corresponding to $10\, \mu$m. Our yeast experiments showed lengths between $8$ to $13\,\mu$m. Here, we investigate the impact of changing chain length on the emergent morphologies and transitions between them.

Fig.~\ref{fig:ShapeN256}$(a)$ shows representative steady-state configurations for four different combinations of $(Pe, \tilde{\epsilon})$ values, illustrating distinct morphologies, namely, SCR, coexistence, open helix, and globule structures at four different chain lengths $N$. Clearly, the morphologies remain unchanged across these different chain lengths, indicating the robustness of our results. 

In Fig.~\ref{fig:ShapeN256}$(b)$, we show the dependence of the size and shape on the activity at $\tilde \e=1$ for an active chain with $N=256$. The results are clearly comparable to that of $N=100$ chain presented in Fig.\ref{fig:shape1}($a$), with even the transition point to open helix at $Pe \approx 0.1$ remaining unchanged. This constitutes clear evidence of chain-length independence of the results obtained for different morphologies and transitions.  

}

\begin{figure*}[t!]
\centering
  \includegraphics[width=17cm]{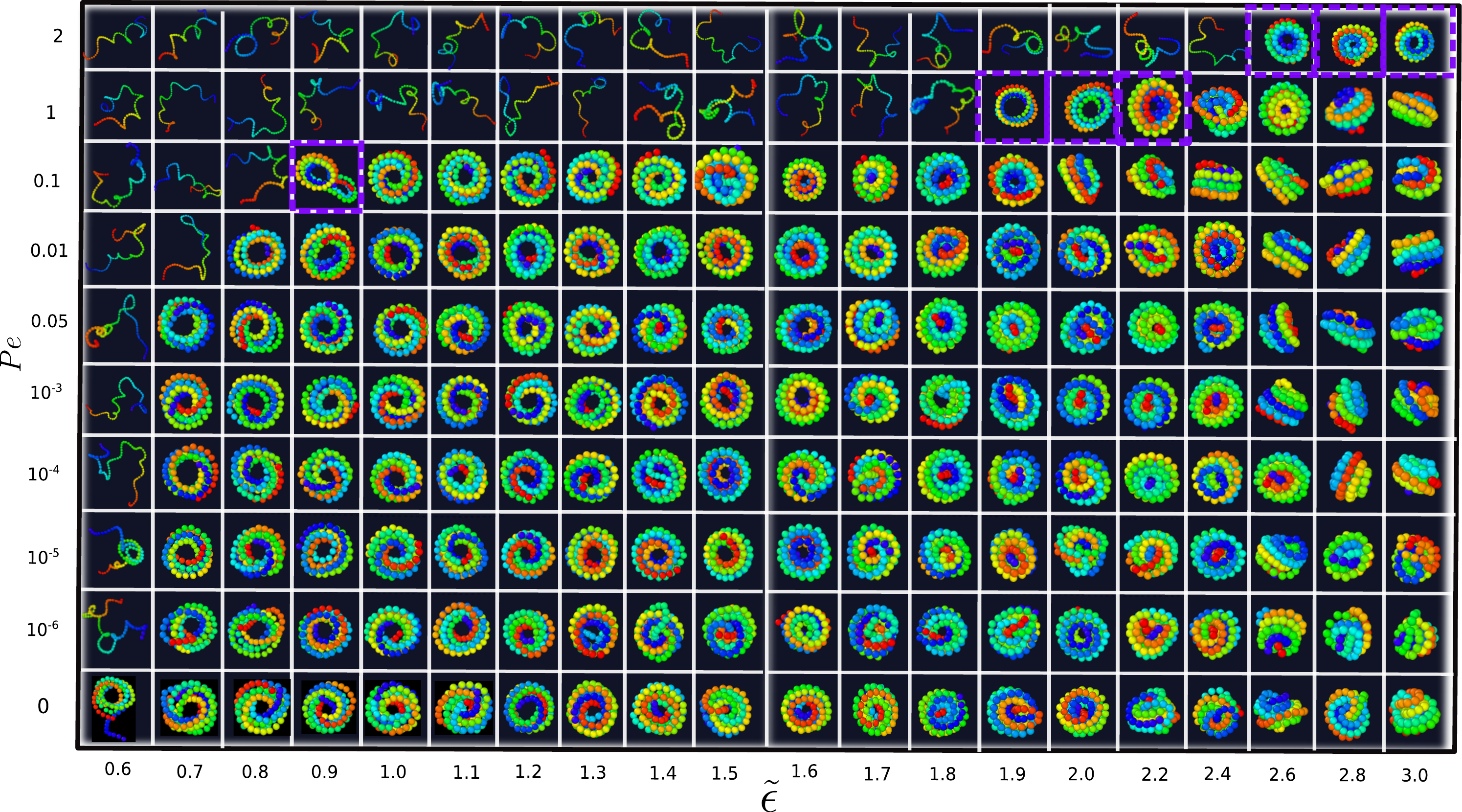}
\caption{Configurational phase diagram in the $Pe-\tilde\e$ plane. The system exhibits various structures across different ranges of activity $Pe$ and lateral attraction strength $\tilde\e$. The chain maintains a stable helical structure for very low values of $\tilde\e$ and forms a ring at intermediate $\tilde\e$ values across a broad range of $Pe$. At higher $\tilde\e$ and lower $Pe$ values, the chain adopts globule structures. Increased $Pe$ tends to favor an open chain conformation by disrupting contacts between chain segments. Violet square boxes denote the coexistence of ring, globule, and open helical structures forming near transition points, where the chain spends maximum time in an open conformation and also forms structures like rings or globules. Note that in terms of another measure of activity defined earlier in Sec.~\ref{sec_model}, $Pe$ ranges from $10^{-2}$ to $2\times 10^4$.
}
\label{fig:configPeEp}
\end{figure*}

\dc{
\section{Outlook}\label{sec:conclusion}
}
Our study addresses the pivotal question of how treadmilling and lateral attraction contribute to the formation of the FtsZ filament's ring structure during cell division. 
Given the crucial dependence of bacterial cell cycle on the Z-ring formation, disruptive mutations in FtsZ are detrimental to cell survival and their effects on polymerization cannot be studied in a cellular {\em melieu}. We circumvent this issue by using fission yeast as a host cell to express the various FtsZ mutants, where the filaments remain freely floating in 3D space and show distinct morphologies such as ring, open helix, and globule. In one specific case with slow treadmilling, we also observed a spooling mechanism of ring formation.}

Using a coarse-grained theoretical model that treats the FtsZ filament bundles as tangentially active chiral semiflexible polymers, we accurately reproduced all these morphological phases observed experimentally. While lateral attraction stabilizes the ring structure, it fails to induce the observed net rotation without treadmilling activity. We obtained a detailed phase diagram exploring ring stability as a function of activity and attraction, revealing a stable ring state over a broad range of treadmilling activity before transitioning into an open helix through phase coexistence, indicative of a first-order phase transition. In fact, treadmilling flattens and widens otherwise globule states to stabilize rings which can generate contractile stresses in their surroundings. The transition from the stable ring state to the open helix state is marked by increased size fluctuations and extensile response. 

\dc{Through independent numerical simulations, we confirmed that the observed morphological phases and transitions between them remain unchanged when replacing the tangentially active chain model with a treadmilling model that conserves filament length. Moreover, the results are robust even with variations in chain length.}
%
%

\dc{Our phase diagram predicts that ring formation arises from an open helix with reduced treadmilling activity, consistent with recent observations of Z-ring formation in bacteria. Bundling of FtsZ protofilaments by cross-linking Zap proteins can reduce GTPase activity~\cite{Hale2011, DurandHeredia2012, Huang2013}, leading to slower treadmilling. According to our model, this reduction, combined with inter-segment attraction, favors ring formation over open chains. This is supported by experiments showing that overexpression of ZapC or ZapD results in a highly condensed, hyper stable Z-ring~\cite{DurandHeredia2011, Hale2011}. Our model suggests that ancestral bacterial cells may have evolved mechanisms to control FtsZ shape across the cell cycle by modulating treadmilling activity that in turn can generate force.

During pre-divisional stages in bacteria, FtsZ forms open filaments that initiate ring formation before division~\cite{Thanedar2004, Walker2020}. The Min system likely evolved to regulate treadmilling by accelerating filament dissolution and preventing premature assembly~\cite{Arumugam2014}. Spatial control of FtsZ assembly could then be mediated by chemical gradients, such as those from the MinCDE system~\cite{Huang&Wingreen2005, Lutkenhaus2007}. Additionally, the forces generated within the ring may drive cytokinesis~\cite{Osawa2013}.  

Moreover, FtsZ has been shown to assemble into rings and toroids in bulk solutions and yeast cytoplasm, independent of membrane anchoring~\cite{Srinivasan2008, Popp2009}. Our predictions of emergent morphologies and their transitions are directly relevant to these yeast experiments. However, correlating treadmilling activity with the observed morphologies in yeast remains a key experimental challenge that could validate our theoretical predictions.
}

\dc{One limitation of our model is that it assumes FtsZ filaments to be free in 3D space, which aligns with our yeast experiments but contrasts with bacterial cells, where, FtsZ filaments are anchored to the inner membrane by proteins like ZipA, FtsA, or SepF, which also assemble into filaments~\cite{Hale1997, Pichoff2005, Duman2013, Pende2021, Radler2024}. Such anchoring could have additional effects on the filament properties. 
FtsZ filaments themselves have a mild curvature with diameters similar to that of a cell~\cite{Srinivasan2008, Loose2014, Radler2022, Erickson2009a}. 
Within a bacterial cell, the dynamic nature of these filaments may play a role in sensing membrane curvature, as has been proposed for MreB~\cite{Ouzounov2016, Hussain2018, Shi2020}.
The membrane anchor protein, FtsA, an actin homolog, forms an intrinsically curved co-polymer with the FtsZ filaments only upon activation by FtsN, favoring the filament (co-polymer) alignment to membrane curvature~\cite{Nierhaus2022}. Our model can be adapted in the future to incorporate co-polymers and membrane anchoring.
 }

While our model treats FtsZ as an assembled polymer, we recognize that the Z-ring's structure in bacteria is more intricate. It forms from multiple shorter treadmilling filaments near a curved membrane surface, merging to create the Z-ring. Models that capture this initial assembly, incorporating treadmilling rates, and account for filament ordering on a curved membrane could offer further insights into bacterial cytokinetic ring formation. Our findings, nonetheless, enhance the understanding of FtsZ ring dynamics and provide a basis for further study into bacterial cytokinesis regulation and the evolutionary link between FtsZ protofilaments and microtubule structures.
\dc{Finally, beyond FtsZ, our model could be extended to explore the origins of cytoskeletal tubes formed by stacking of multiple rings, as in archaeal Odin.tubulin~\cite{Akil2022}.}

\vskip .4cm

\section{ Acknowledgement}\label{sec:5}
The numerical simulations were performed using 
SAMKHYA, the High-Performance Computing Facility
provided by the Institute of Physics, Bhubaneswar. Image credits: Authors acknowledge Ajay Kumar Sharma and Sakshi Mahesh Poddar for FtsZ images and Mirza Salim Beg at the Center for Interdisciplinary Sciences at NISER, Jatni. We thank Amir Shee and Subhashree S Khuntia for the FtsZ persistence length analysis. DC acknowledges a research grant from the Department of Atomic Energy (DAE), India (1603/2/2020/IoP/R\&D-II/150288) and thanks the International Centre for Theoretical Sciences (ICTS-TIFR), Bangalore, for an Associateship. RS acknowledges funding support from NISER, DAE and a research grant from the Department of Biotechnology, DBT, India (BT/PR15183/BRB/10/1443/2015). \dc{This research was supported in part by ICTS-TIFR during a visit for participating (RS and DC) in the program Dynamics of Complex Systems - 2017 (ICTS/Prog-dcs/2017/05).}

\appendix

\begin{figure}[t!]
\centering
  \includegraphics[width=6cm]{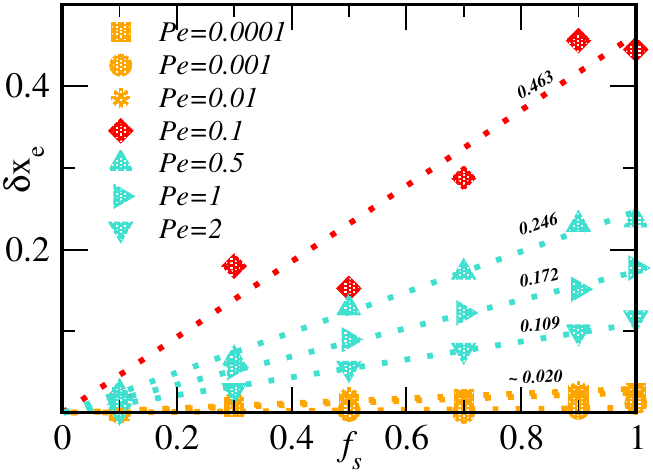}
\caption{Extension $\delta x_{e}$ as a function of external force $f_s$ for various values of $Pe$ and fixed $\tilde \epsilon=1$. The data points are color-coded: orange for the ring phase, turquoise for the open helix conformations, and red indicating the transition point ($Pe=0.1$). Dotted lines depict linear fits to the data for various $Pe$ with their respective slope values. }
\label{fig:ResFunc}
\end{figure}

\section{Persistence length of open chains}\label{app_Llp}
We used movies of the FtsZ mutant EcFtsZ Q47KD86K that remains in open chain conformation to extract effective persistence length $\ell_p$ by evaluating tangent-tangent correlation and fitting with $\langle \hat u(s)\cdot \hat u(0)\rangle=\exp(-s/\ell_p)$ over shorter length scales. We used FIJI to extract the chain conformation on which we performed the analysis. The correlation over longer separations is known to deviate from exponential to power-law decay due to, e.g., self-avoidance. 

\begin{figure}[t!]
\centering
  \includegraphics[width=8cm]{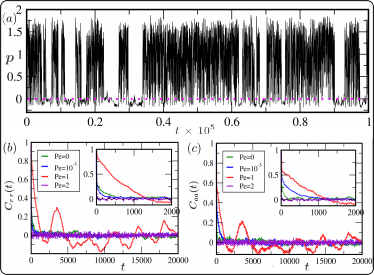}
\caption{$(a)$ Full time series of prolateness $p$ at $Pe=1$ and $\tilde \epsilon =2$ where ring (marked with $R1$, $R2$), globule ($G$) and open helix ($OH$) co-exist. The dashed magenta line corresponds to $S=0$. $(b)$,\,$(c)$ are two-time auto-correlation functions of radius of gyration ($r_g$) and asphericity ($ \mathfrak{a}$), respectively, for various $Pe$ values at $\tilde \epsilon =2$.
}
\label{fig_CorrS}
\end{figure}

 \begin{figure}[b!]
\centering
  \includegraphics[width=7cm]{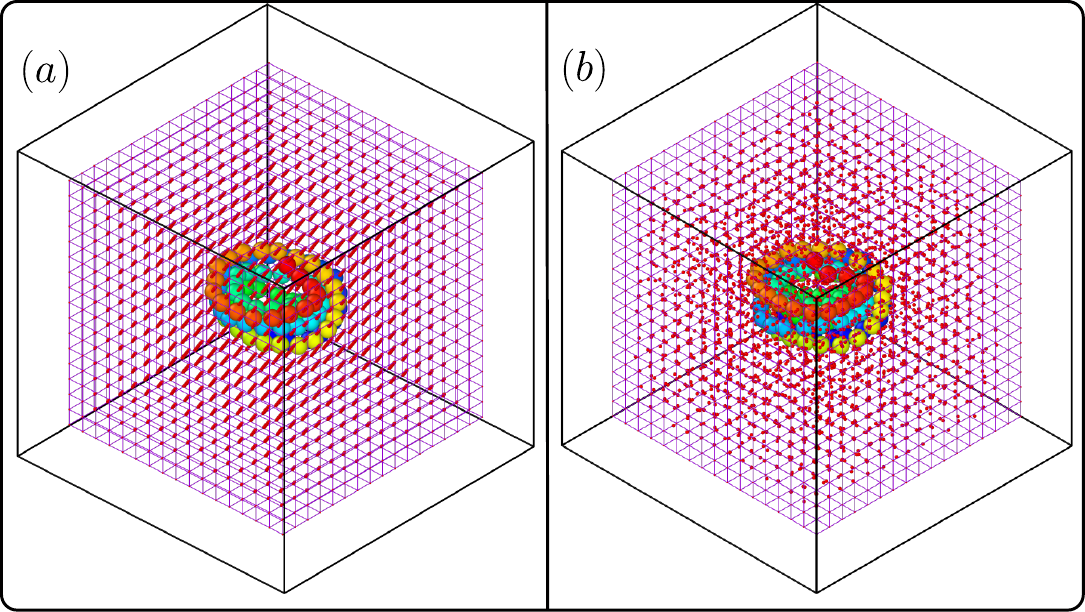}
\caption{$(a)$ The cubic lattice embedding the active ring. The boundary points (small red-colored spheres) at the six faces remain fixed, while the rest of the lattice can distort. The monomers of the chain are bonded with the lattice points within a cutoff distance $\sigma$. $(b)$~A typical configuration with distortion of the lattice at $Pe=1.$  
The bounding box, indicated by black lines, is solely for visual representation.
}
\label{fig:gridwall}
\end{figure}

\begin{table}[h!]
    \centering
    \begin{tabular}{|c |c |c |c| c|} 
    \hline
         data sets & $L$ & $\ell_p$ & $\frac{\Delta \ell_p}{\ell_p}$ & $L/\ell_p$\\
         \hline
         1 & 250 &63.39 & 3.93\% & 3.94\\
         2 & 290 &60.84 & 2.57\% & 4.76\\
         3 & 376 &65.26 & 2.42\% & 5.76\\
         \hline
    \end{tabular}
    \caption{The values of $L$ and $\ell_p$ listed in the table are in units of pixels where each pixel has a size $0.0346\,\mu$m. The length of the chains are $L \approx 10 \, \mu$m, and vary between $8\,\mu$m to $13\,\mu$m. The effective persistence length $\ell_p \approx 2\,\mu$m, equivalent to wild type FtsZ values measured earlier~\cite{Srinivasan2008}.}
    \label{tab:my_label}
\end{table}

\section{Configurational phase diagram}\label{appendA}
The main text presents a configurational phase diagram for a few selected parameter values in the $Pe-\tilde \epsilon$ plane. In Fig.~\ref{fig:configPeEp}, we show a full conformational phase diagram over a wider range of parameter values.

\section{Extensile response}\label{appendB}
In Fig.~\ref{fig:ResFunc}, we show the change in the end-to-end distance along the $x$-direction with respect to the steady state distance  $\delta x_{e}$ ($:= \langle x_{e} \rangle_{f_s} - \langle x_{e} \rangle_{f_s=0}$) as a function of the applied stretching force $f_s$ at different $Pe$ and fixed $\tilde \epsilon=1$. For all $Pe$ values, $\delta x_{e}$ scale linearly with small $f_s$ as shown with the dotted lines in Fig.~\ref{fig:ResFunc}. The linear fits are used to calculate the extensile linear response $\chi=\f{\p \d x_e}{\d f_s}$ used in the main text. Note that because of self-avoidance, compression of the rings that already have a compact structure is strongly resisted, and the slope of the response sharply changes for negative $f_s$.   

  \begin{figure}[t!] 
\centering
  \includegraphics[width=8cm]{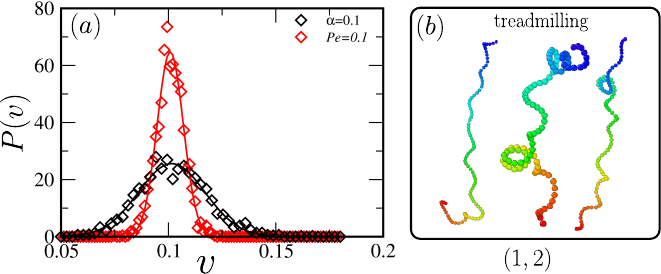}
\caption{\dc{
$(a)$~Comparison of steady-state velocity distributions for tangentially active chains with $Pe=0.1$ (red) and treadmilling filaments (black) with turnover rate $\alpha = 0.1$. Points represent simulation results, and lines are Gaussian fits. ($b$)~Representative steady-state configurations at $\alpha = 1$ and $\tilde{\epsilon} = 2$ from the treadmilling model.
}
}
\label{fig:ShapeN100Tm}
\end{figure}

\section{Dynamics: Time-series and correlation}
\label{append_tseries}
Here, we show a long-time series of prolateness $-1/4 \leq p \leq 2$ near a transition point $\tilde \e=2$ and $Pe=1$. The value of $p$ shows large excursions between open helix (large positive $p$), globule $p=0$, and ring $p<0$ states; see Fig.\ref{fig_CorrS}($a$).   
The correlation functions $C_{rr}(t)=\la \d r_g(t) \d r_g(0)\ra/\la \d r_g^2(0) \ra$ for the radius of gyration $r_g(t)$ and $C_{\mathfrak{a} \mathfrak{a}}(t)=\la \d \mathfrak{a}(t) \d \mathfrak{a}(0)\ra/\la \d \mathfrak{a}^2(0) \ra$ asphericity $ \mathfrak{a}(t)$ show similar features as that of $C_{pp}(t)$ shown in the main text. The apparent oscillations in the correlations at $Pe=1$ do not correspond to any specific oscillatory Fourier mode, as verified using the corresponding power spectra having features similar to $S_{pp}(\nu)$ shown in the main text.

\section{Distortion of embedding matrix}\label{appendD}
Fig.~\ref{fig:gridwall}$(a)$ shows the initial cubic lattice with the active ring embedded in it. Fig.~\ref{fig:gridwall}$(b)$ shows a typical steady-state configuration with the distortion of the lattice points and stretching of the bonds at $Pe=1$.

 \section{Comparison with treadmilling model}\label{appendTreadmil}

We first consider the velocity distribution of an active bead-spring filament in one dimension (1D) described by the bond energy $U_b$ as in Sec.\ref{sec_model} and self-propulsion quantified by the P{\' e}clet number $Pe$. In the presence of translational diffusion, as in Eq.\eqref{eq_lange}, the distribution shows a Gaussian nature. The numerical simulation results with $Pe=0.1$ and a fit to the Gaussian  $f(v) = \frac{1}{\sqrt{2\pi\mathcal{A}}} \exp\left(-\frac{(v-\bar v)^2}{2\mathcal{A}}\right)$ giving mean $ \bar v \approx 0.1$ and variance $\mathcal{A}=3.71 \times 10^{-5} $ is shown in Fig.\ref{fig:ShapeN100Tm}($a$). The related stochastic treadmilling filament model with a turnover rate $\a$ in which monomers from the tail are removed and inserted at the head has a fixed length and mean velocity $\a r_0$. The mapping completes with $Pe= \a r_0^2/D$. The simulation results from this model at $\a=0.1$ show a broader velocity distribution and fits with Gaussian with the same mean velocity $\bar v \approx 0.1$ but larger variance $\mathcal{A}=2.4\times 10^{-4}$; see Fig.\ref{fig:ShapeN100Tm}($a$).

In Figure \ref{fig:ShapeN100Tm}$(b)$ we show representative steady-state configurations in three dimensions for the treadmilling model at $\alpha = 1$ and $\tilde{\epsilon} = 2$. At this parameter value, which marks the transition boundary in the active filament model, the treadmilling filament adopts an open helix morphology, with transient local rings forming along its length, running from one end to the other.


%
\end{document}